\documentclass[a4paper,12pt]{article}

\usepackage{amstext,amsmath,amssymb,amsthm}  

\usepackage[normalem]{ulem}  
\usepackage{graphicx}  
\usepackage{booktabs}  
\usepackage{bm} 

\usepackage[comma,sort,authoryear]{natbib}
\usepackage{hypernat}  

\usepackage{vmargin}  
\usepackage{authblk}

\usepackage{xcolor}
\definecolor{dark-red}{rgb}{0.8,0.15,0.15}
\definecolor{dark-blue}{rgb}{0.15,0.15,0.6}
\definecolor{medium-blue}{rgb}{0,0,0.8}

\usepackage[pdftex,breaklinks,colorlinks]{hyperref}
\hypersetup{  
  pdftitle =
    {Heating the coffee by looking at it. Or why quantum measurements are physical processes},
  linkcolor={dark-red},
  citecolor={dark-blue},
  urlcolor={medium-blue}
}

\begin{document}

\numberwithin{equation}{section}
\numberwithin{figure}{section}
\allowdisplaybreaks[1]  

\title{Heating the coffee by looking at it. Or why quantum measurements are physical processes}

\author[,1,2,3,4,5]{Pablo Echenique-Robba\footnote{{\footnotesize \
\href{mailto:pablo.echenique.robba@gmail.com}{\texttt{pablo.echenique.robba@gmail.com}} ---
\href{http://www.pabloecheniquerobba.com}{\texttt{http://www.pabloecheniquerobba.com}}}}}

\affil[1]{Instituto de Qu\'{\i}mica F\'{\i}sica Rocasolano, CSIC, Madrid, Spain}
\affil[2]{Instituto de Biocomputaci\'on y F{\'{\i}}sica de Sistemas Complejos (BIFI), Universidad de Zaragoza, Spain}
\affil[3]{Zaragoza Scientific Center for Advanced Modeling (ZCAM), Universidad de Zaragoza, Spain}
\affil[4]{Departamento de F{\'{\i}}sica Te\'orica, Universidad de Zaragoza, Spain}
\affil[5]{Unidad Asociada IQFR-BIFI, Madrid-Zaragoza, Spain}

\date{\today}

\maketitle

\begin{abstract}

Using a very simple \emph{Gedankenexperiment}, I remind the reader that
(contrary to what happens in classical mechanics) the energy of a quantum
system is \emph{inevitably} increased just by performing (some) textbook
measurements on it. As a direct conclusion, this means that some measurements
require the expenditure of a finite amount of energy to be carried out. I also
argue that this makes it very difficult to regard measurements as disembodied,
immaterial, informational operations, and it forces us to look at them as
physical processes just like any other one.
\vspace{0.4cm}
\vspace{0.2cm}\\

\end{abstract}

\section{Introduction}
\label{sec:introduction}

In everyday life, we know that a cold coffee mug does not heat up simply by
looking at it. We have to interact in some stronger way with it than just
looking if we want to drink the coffee hot; for example, we have to put it in
the microwave. Similarly, we know that no other type of energy can be
transferred to a physical system \emph{just by finding out something about it}:
watching at the blinking dot in a radar screen does not change the speed of the
corresponding plane, looking at Mars through the telescope does not alter the
weather there, and watching the football fly in the TV does not increase the
probability that it ends up scoring a goal---no matter how much the fans would
like it (or even assume it).

Classical mechanics underwrites this intuition---this ``folk
physics''---formally and mathematically. In classical mechanics it is tacitly
assumed that the possible disturbance on the system that may be produced by
measuring any of its physical properties can be in principle reduced to naught.
That is, classical mechanics does admit the possibility that measurements cause
a non-negligible disturbance (for example, if you measure the position of a
football by throwing tennis balls at it and looking which ones bounce back at
you), but nothing in the formalism precludes you from designing better and
better measuring devices until you make the disturbance vanishingly small.
Famously, this is \emph{not} the case in quantum mechanics; as I will try to
show here in a simple but dramatical example.

What happens \emph{exactly} when we perform a measurement on a quantum system
is still an open question, but one thing is clear: it is not at all like the
classical case. The literature of quantum foundations is a bit messy at the
moment with many different accounts competing for prime time in the journals
\citep{Echenique-Robba2013}, and most of them differ in the characterization of
quantum measurements---a central piece of the puzzle. Specifically, there are
at least two points of view from which one can approach measurements: the
\emph{informational} and the \emph{physical}. The first one emphasizes that
measurements are ways that we humans have to \emph{know things about} the
quantum system, to extract information from it in the form of measurement
outcomes. The physical point of view, in turn, focuses in the notion that a
quantum measurement is a \emph{physical process} just like any other one; even
if possibly belonging to a given class with some specific properties. For
example, that it must involve at least the system and a macroscopic measuring
device, or that the latter must display well defined, distinguishable and
meaningful outcomes when the measurement is over.

In this short note, I will present a very simple (almost trivial)
\emph{Gedankenexperiment}\footnote{\label{foot:gedanken} The word is German,
and that's why it begins with a capital letter, like all good German nouns. It
means ``thought experiment''. That is, a very idealized version of an
experiment that you only carry out in your imagination, but that you assume to
be nevertheless doable in the lab. You also assume that the ideal conditions
might never be strictly attained in reality, but approaching them as much as
one wants is only a matter of technical ability. Although ``Gedanken'' is the
plural noun meaning ``thoughts'', it functions as an adjective here.} that
seeks to emphasize this latter, physical point of view. My only aim is to show
as clearly and as simply as possible that regarding measurements as physical
processes is \emph{inescapable}. In particular, this implies that any approach
to quantum measurements might involve a combination of informational and
physical considerations, but it could never be constituted by the first type
\emph{alone}. The reader may feel that everybody must agree---or even that
everybody \emph{does} agree---with this thesis, but, given the polyphonic
nature of the positions and debates in the field, I think it is better to be
safe than sorry.

Also, I will make my point in a very concise way; something which might help to
nail down and more fruitfully confront any statement related to the
informational vs.~physical dichotomy. This extreme simplicity together with the
idiosyncratic nuances of my personal approach to the topic are possibly the
only original ingredients that the reader will find here. I hope they are
enough to make the ride interesting and useful, but if they are not, don't
worry. The ride is very short anyway.

\section{A simple \emph{Gedankenexperiment}}
\label{sec:gedanken}

Let us consider a spin-1/2 quantum system with a Hamiltonian operator which is
proportional to the $z$-component of the spin:
\begin{equation}
\label{eq:H}
\hat{H} = - \alpha \hat{S}_z = - \frac{\alpha \hbar}{2} \hat{\sigma}_z \ ,
\end{equation}
where we have defined the spin-$z$ operator by:
\begin{equation}
\label{eq:S_z}
\hat{S_z} := \frac{\hbar}{2}
\hat{\sigma_z} := \frac{\hbar}{2}
\left(
\begin{array}{cc}
1 & 0 \\
0 & -1 \\
\end{array}
\right) .
\end{equation}

Now imagine that we are able to prepare, at time $t_0 = 0$, $N$ copies
of the system in the eigenstate $| z+ \rangle$ of $\hat{S}_z$ corresponding to
the positive eigenvalue $+\hbar/2$. This state is obviously also the ground
state of $\hat{H}$ with eigenvalue (energy) equal to $-\alpha\hbar/2$.
Therefore, it will evolve in time with just a phase:
\begin{equation}
\label{eq:zplus_t}
| \psi(t) \rangle =
 \exp \left( - i \frac{t}{\hbar} \hat{H} \right) | \psi(0) \rangle =
 \exp \left( i \frac{\alpha t}{2} \hat{\sigma}_z \right)
  | z+ \rangle = \exp \left( i \frac{\alpha t}{2} \right) | z+ \rangle \ .
\end{equation}
That is, it will remain an energy eigenstate corresponding to the minimum energy
$-\alpha\hbar/2$ until we do something
else.\footnote{\label{foot:normalized_rays} I am aware that quantum states are
not represented by elements of a Hilbert space, but by the corresponding rays,
i.e., by the corresponding 1-dimensional linear subspaces spanned by them. To go
from rays to vectors (kets), one has to fix the norm and an arbitrary phase. In 
this paragraph and in everything that follows, all kets are assumed normalized 
and the phase is chosen to make the expressions as simple as possible.}

Next (at some time $t_1$ after preparation), let us measure
\begin{equation}
\label{eq:S_x}
\hat{S_x} := \frac{\hbar}{2}
\hat{\sigma_x} := \frac{\hbar}{2}
\left(
\begin{array}{cc}
0 & 1 \\
1 & 0 \\
\end{array}
\right)
\end{equation}
on each and every prepared copy. According to the accepted understanding of 
quantum measurement in textbooks, the result of such an action is that the 
state $| \psi(t_1) \rangle$ will collapse instantaneously to any one of the two
eigenstates $\{| x- \rangle, | x+ \rangle\}$ of $\hat{S}_x$ with probabilities
$\{P(x-;t_1), P(x+;t_1)\}$ respectively. By Born's rule, we know that:
\begin{equation}
\label{eq:prob_x_1}
P(x\pm;t_1) = \big| \langle x\pm | \psi(t_1) \rangle \big|^2 =
 \bigg| \exp \left( i \frac{\alpha t_1}{2} \right) \langle x\pm | z+ \rangle
 \bigg|^2 = \big| \langle x\pm | z+ \rangle \big|^2 \ .
\end{equation}
Since
\begin{equation}
\label{eq:x_kets}
| x \pm \rangle = \frac{1}{\sqrt{2}} \left( | z- \rangle \pm
  | z+ \rangle \right) \ ,
\end{equation}
and of course $\langle z- | z+ \rangle = 0$, and $\langle z- | z- \rangle = 
\langle z+ | z+ \rangle = 1$, we have that:
\begin{equation}
\label{eq:prob_x_2}
P(x\pm;t_1) = \frac{1}{2} \ .
\end{equation}
That is, half of the times the $\hat{S}_x$ measurement will collapse the state
onto $| x- \rangle$, the other half it will collapse it onto $| x+ \rangle$, and
the probabilities are time independent.

The subsequent evolution of any of these eigenstates of $\hat{S}_x$ can be
easily obtained using eq.~(\ref{eq:x_kets}) and the fact that the eigenstates
$\{| z- \rangle, | z+ \rangle\}$ of $\hat{S}_z$ are also eigenstates of 
$\hat{H}$ with eigenvalues $\{\alpha\hbar/2,-\alpha\hbar/2\}$ respectively:
\begin{equation}
\label{eq:xpm_t}
| \psi(t>t_1) \rangle =
\frac{1}{\sqrt{2}} \left[
  \exp \left( -i \frac{\alpha (t-t_1)}{2} \right) | z- \rangle \pm
  \exp \left( i \frac{\alpha (t-t_1)}{2} \right) | z+ \rangle \right] \ .
\end{equation}

The last step in the \emph{Gedankenexperiment} is to perform an additional
measurement, now of $\hat{S}_z$, at a time $t_2 > t_1$. Orthodox presentations
of quantum mechanics indicate again that this will produce the collapse of the
state $| \psi(t_2) \rangle$ onto either $| z- \rangle$ or $| z+ \rangle$. The
corresponding Born probabilities are also time independent, and they can be 
easily computed:
\begin{equation}
\label{eq:prob_z}
P(z\pm;t_2) = \frac{1}{2} \ .
\end{equation}
That is, half of the systems will end up in state $| z- \rangle$ after the
$\hat{S}_z$ measurement, and the other half will end up in state $| z+ \rangle$.

But wait! States $\{| z- \rangle, | z+ \rangle\}$ are also eigenstates of the
Hamiltonian operator, which means that any energy measurement at $t_3 > t_2$ is
now certain to produce $E = \alpha\hbar / 2$ for the first one of them, or 
alternatively $E = -\alpha\hbar / 2$ for the second. This in turn means that, 
if we produced $N$ copies of our quantum system in eigenstate $| z+ \rangle$ at 
$t_0 = 0$, we began with a situation in which the total energy was
\begin{equation}
\label{eq:E_tot_t0}
E_\mathrm{tot}(0) = - N \frac{\alpha\hbar}{2} \ ,
\end{equation}
and we ended up with a situation in which
\begin{equation}
\label{eq:E_tot_t3}
E_\mathrm{tot}(t_3) = \frac{N}{2} \frac{\alpha\hbar}{2}
 - \frac{N}{2} \frac{\alpha\hbar}{2} = 0 \ .
\end{equation}
That is, we increased the energy of our (collection of) systems by 
$N\alpha\hbar / 2$. Or, if you want to put it more dramatically, \emph{we 
heated the coffee by just looking at it} (twice).

\section{From \emph{Gedanken} to \emph{Laboratorium}}
\label{sec:laboratorium}

The previous \emph{Gedankenexperiment} is ``gedanken'' but doable. One possible
way of maybe realizing it in a laboratory is through the famous Stern-Gerlach
setup [which you can look up in many places, but I have looked up in
\citep{Peres2002}].

The basic idea is to use some kind of source of neutral particles with spin,
generate a collimated beam, and make it go through a region in which there is
an inhomogeneous magnetic field; normally produced by a suitable magnet.
Combined with some sort of detector after the magnet, this constitutes an
experimental embodiment of a quantum ``spin measurement''. Indeed, if the
mentioned detector is (say) some kind of screen which can be excited anywhere
on its 2-dimensional surface, we find that only two discrete spots appear; one
corresponding to the eigenstate of the spin operator in the direction of the
magnetic field, the other corresponding to the orthogonal one, i.e., to the
eigenstate associated to the opposite direction. Since the classical behavior
that one would expect (assuming that spin is some kind of magnetic dipole)
is to obtain a continuous band between the two spots corresponding to all
possible intermediate orientations between completely aligned with the magnetic
field and completely anti-aligned to it, the result of this basic Stern-Gerlach
setup is typically regarded as demonstrating that spin is \emph{quantized},
i.e., that spin is a quantum property; not a classical one.

This is nice and correct, but I don't want to prove that quantum mechanics is 
necessary here. I want to show that, \emph{assuming it is}, we can use it to 
heat coffee just by looking. So let us use the Stern-Gerlach idea to produce a
more complicated---but still sketchy---setup that can move the 
\emph{Gedankenexperiment} in the previous section closer to a real 
\emph{Laboratoriumexperiment}.\footnote{\label{foot:laboratorium} You don't need
to know German to understand this one.}

\begin{figure}[!t]
\begin{center}
\includegraphics[scale=0.3]{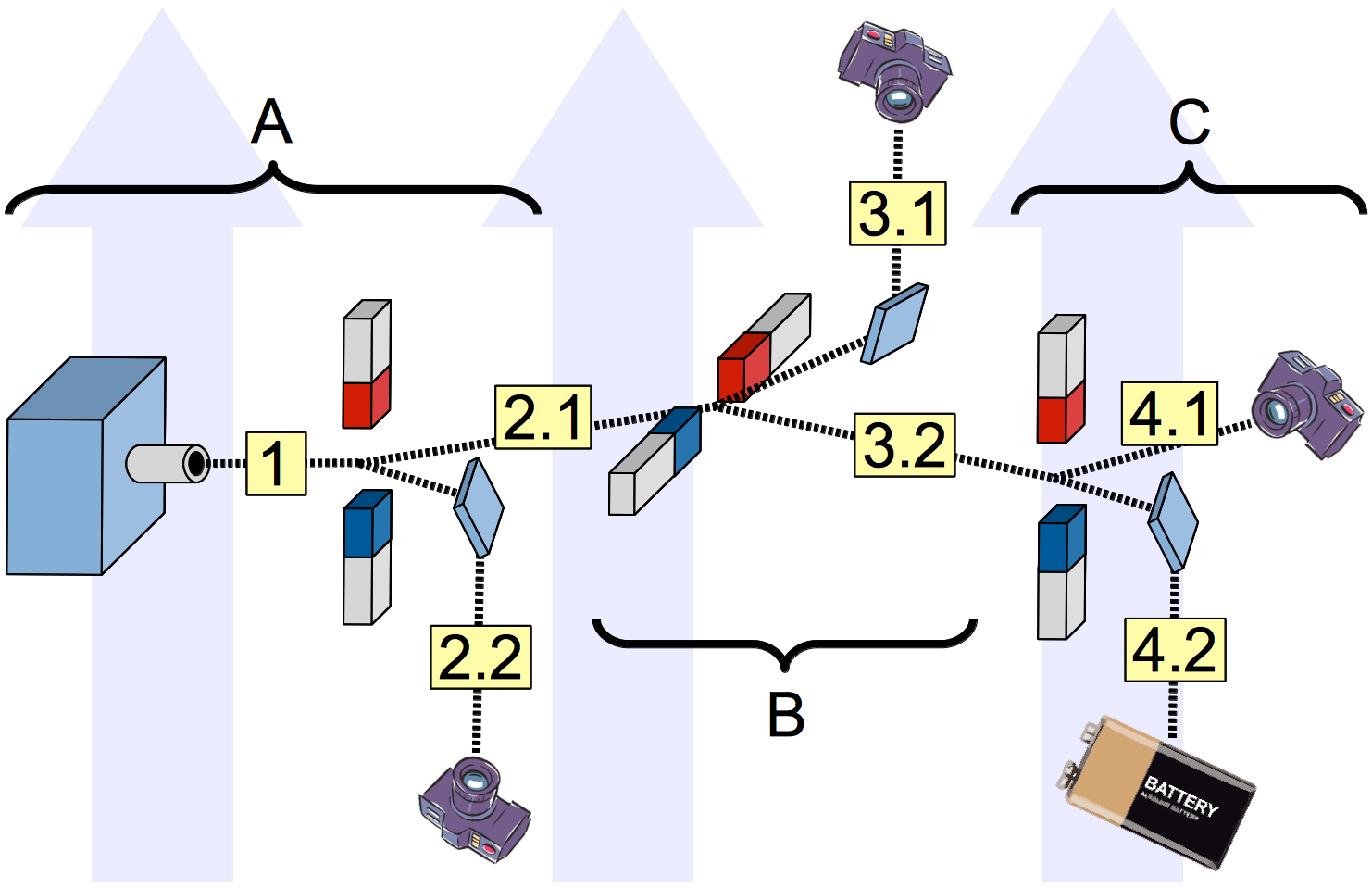}
\caption{\label{fig:stern-gerlach}{\small Cartoon depiction of the
\emph{Laboratoriumexperiment} described in detail in 
sec.~\ref{sec:laboratorium} and constituting a possible physical embodiment of 
the \emph{Gedankenexperiment} in sec.~\ref{sec:gedanken}.}}
\end{center}
\end{figure}

A cartoon scheme of the whole thing can be found in
fig.~\ref{fig:stern-gerlach}: On the leftmost part, we have some kind of source
that emits our spin-1/2 particles in a completely random state; i.e., in the
(statistical) state described by a density matrix $\hat{\rho}$ equal to the $2
\times 2$ identity matrix $\hat{I}$. The light-blue arrows in the background
represent an homogenous magnetic field in the positive $z$ direction which is
assumed to be present at every point of the setup, and which explains why
$\hat{H}$ in eq.~(\ref{eq:H}) is proportional to $-\hat{S_z}$. The beam that
comes out of the source is labeled as `1' and it is depicted as a discontinuous
line; like the rest of beams in the figure.\footnote{\label{foot:beams} Of
course, quantum mechanics (via the uncertainty principle) banishes well defined
trajectories out of existence. As a consequence, all the ``beams'' I have drawn
must be understood as ``semiclassical metaphors'' which are only intended to
help us reason comfortably, and which are in any case \emph{harmless}. Since we
materialize the ``splitting of beams'' at each step by performing a quantum
measurement with definite---even if unpredictable---outcomes, all the ``beam
talk'' is in fact consistent and one could dispose of it if extreme purism was
intended.} Beam 1 is assumed to consist of $2N$ particles all of them prepared
identically, and we can imagine that the intensity is so low that the particles
come out of the source one by one.

Just after the source, we have our first Stern-Gerlach magnet oriented along
the $z$ axis. Instead of making the resulting beams impinge on a screen (which
would yield the two proverbial spots but would prevent further actions on the
particles), we place a mirror at the location where the spot corresponding to
$| z- \rangle$ would appear, and a detector (depicted as a camera) just after
it. The device formed by the magnet, the mirror and the camera is a physical
embodiment of a $\hat{S}_z$ measurement, with the particularity that the
systems with spin $z+$ are not ``destroyed'' but move on. If we add the source
to this device, we can regard the whole sub-setup labeled by `A' as a
``preparation'' of the system onto the state $| z+ \rangle$. That is, the state
of every particle in beam 2.1 is $| z+ \rangle$ and its energy is the minimal
one, $-\alpha\hbar/2$ (rigorously speaking, an energy measurement on any
particle in beam 2.1 is certain to yield this value). Beam 2.1 can thus be seen
as the starting point of the \emph{Gedankenexperiment} in
sec.~\ref{sec:gedanken}, and the total energy it contains is $-N\alpha\hbar/2$,
since only half of the original particles come out of the preparing device A. In
tab.~\ref{tab:stern-gerlach}, the number of particles, the states and total
energies of each beam are summarized for quick reference.

\begin{table}[!t]
\begin{center}
\begin{tabular}{l@{\hspace{20pt}}l@{\hspace{20pt}}l@{\hspace{20pt}}l}
\toprule
beam & \#particles & state & $E_\mathrm{tot}$ \\
\midrule
1   & $2N$
    & $\hat{\rho} = \hat{I}$
    & $0$ \\
\midrule
2.1 & $N$
    & $| z+ \rangle$
    & $-N\alpha\hbar/2$ \\
2.2 & $N$
    & $| z- \rangle$
    & $N\alpha\hbar/2$ \\
\midrule
3.1 & $N/2$
    & $| x+ \rangle$
    & $0$ \\
3.2 & $N/2$
    & $| x- \rangle$
    & $0$ \\
\midrule
4.1 & $N/4$
    & $| z+ \rangle$
    & $-N\alpha\hbar/8$ \\
4.2 & $N/4$
    & $| z- \rangle$
    & $N\alpha\hbar/8$ \\
\bottomrule
\end{tabular}
\caption{\label{tab:stern-gerlach}{\small Number of particles, states and total
energies of each beam in the experiment discussed in 
sec.~\ref{sec:laboratorium} and depicted in fig.~\ref{fig:stern-gerlach}.}}
\end{center}
\end{table}

The next piece of equipment (device B) is constituted first by another magnet,
this time oriented along the $x$ axis, which splits beam 2.1 into two new
beams: beam 3.1, which is deflected by a mirror and detected by a camera, and
beam 3.2, which is allowed to proceed to the last part of the setup. Device B
implements a $\hat{S}_x$ measurement, and we assume that we have arranged it in
such a way that the state of every particle in beam 3.1 is represented by $| x+
\rangle$ and those in beam 3.2 by $| x- \rangle$ (but, in fact, nothing
essential changes if we do it the other way around). According to the
calculations in sec.~\ref{sec:gedanken}, the total energy in both beams 3.1 and
3.2 is zero. This already allows us to conclude that, starting from beam 2.1,
which had a total energy of $-N\alpha\hbar/2$, we have increased the energy of
our collection of particles just by measuring their spin along the $x$ axis.
However, in this \emph{Laboratoriumexperiment} I have decided to make this
official by performing a final energy measurement.

Device C is just the same as device B, only that it is oriented along the $z$
axis and the particles that come out of the magnet with state $| z- \rangle$
[half of them, as we know through eq.~(\ref{eq:prob_z})] are not allowed to 
move on, but they are ``stored'' in some sort of ``battery''. Those that come
out with state $| z+ \rangle$ are simply ``detected''.

We can now conclude that, starting from a beam (2.1) with $N$ particles all of
them in their ground state, we have managed to produce $N/4$ particles in an
excited state. We have stored them in a battery and, provided $N$ is large
enough, we may even use them to heat up our coffee. It seems that this setup is
producing less energy than the \emph{gedanken} case in sec.~\ref{sec:gedanken},
but that's only because we have decided to just ``detect'' beam 3.1. If we
channeled this beam to another device of type C, we would then obtain another
$N/4$ particles each one in the excited state of energy $\alpha\hbar/2$, and
the real efficiency of our ``power station'' would attain its theoretical
maximum again.

\section{Discussion and conclusions}
\label{sec:discussion_and_conclusions}

In the previous two sections I have shown how, using the textbook rules for
quantum measurements [see, e.g., \citep{Cohen-Tannoudji1977}], we can transform
$N$ spin-1/2 particles in their lowest energy state into $N/2$ particles in the
same ground state plus $N/2$ particles in a \emph{higher} energy state. I have 
shown that it is very easy to do so, and, what is maybe surprising, that we can 
do it just by performing spin measurements; not by pumping energy into our 
system in some obvious way.

That quantum measurements ``disturb the system'' is a widely known adagio, but I
think that explicitly showing that this disturbance can take the form of 
\emph{a net increase in the system's energy} is a suggestive way of emphasizing 
that a quantum measurement might be a very complicated thing, but it must be a
very complicated thing of an unavoidable physical nature after all. Otherwise, 
how could we inject energy onto the system if the only thing we do is 
measure its properties?

I have chosen energy and not any other magnitude because it seems to me that 
energy has something ``very real'' about it, something ``very tangible''. After 
all, we use it to move our cars and planes, and to light and heat our homes. It 
even has a monetary value! This is all, of course, very informal talk; but not
completely so. This line of though, for example, suggests that---beyond all the
quantum conceptual complications---if energy is being injected into the system,
we must be providing it \emph{at some point}. Otherwise, I have just invented a 
machine that can produce dollars out of thin air and this paper is much more 
important than I first thought.

In the particular setup described in sec.~\ref{sec:laboratorium}, one might
guess that device B, i.e., the part of the experiment that implements the
measurement of spin in the $x$ direction, must play some role in increasing the
energy of the system. Indeed, if we remove it and leave everything else
unchanged, we will find that no particle arrives to the ``battery''. All
$N$ of them will be detected in beam 4.1 as having state $| z+ \rangle$ and
ground-state energy $-\alpha\hbar/2$; i.e., nothing will have changed with
respect to the original beam 2.1. The total energy will also be the same: the
minimum one we started with, $-N\alpha\hbar/2$. It thus seems obvious that
device B must be injecting at least $\alpha\hbar/4$ joules into the system, and
therefore it must be taking this energy from somewhere else. That is, it seems
that we must provide at least $\alpha\hbar/4$ joules from the outside to device 
B if we want that it does its job properly as specified (as if it needs this 
energy to ``rotate'' the spins).

One might want to argue that perhaps this energy expenditure and the system's
energy increase are not necessary features of the measurement process, but 
contingent on the specific way in which the \emph{Laboratoriumexperiment} in
sec.~\ref{sec:laboratorium} has been designed. It is absolutely expected---one
might say---that energy is increased; after all, we are not just ``measuring
spin $x$'', we are doing so with a big fat magnet which we need to power 
somehow. However, the fact that I have established the result in 
sec.~\ref{sec:gedanken} in a completely general way using only the textbook 
rules of quantum measurements allows us to quickly dismiss this objection, and
to confidently bet our money on the truth of the following general constraint:

\begin{quote}
\emph{If you have a spin-1/2 quantum system that has been prepared onto the
``spin-$z$ up'' eigenstate $| z+ \rangle$ of $\hat{S}_z$ and its Hamiltonian is
given by $\hat{H} = -\alpha \hat{S}_z$, then it is \emph{impossible} to measure
the spin of the system in the $x$ direction without expending at least
$\alpha\hbar/2$ joules.}
\end{quote}

One can also complain that, in the particular setup in
sec.~\ref{sec:laboratorium}, the total energy of the original beam 1 produced
by the source is actually zero; hence, no energy is actually being introduced
into the system. If we consider our initial system as beam 1, this is indeed
so. However, and as it is clear from the discussion of both the \emph{gedanken}
and \emph{laboratorium} experiments, this is \emph{not} so if we regard beam
2.1 as our initial system. I cannot find any argument that forbids this choice
to be made, so I must temporarily dismiss the objection, even if it looks
better than the previous one a priori.\footnote{\label{foot:alpha_centauri} It
helps me to imagine that device A might be located in Alpha Centauri, and we
receive beam 2.1 after it has traveled 4.367 light years. Only then we increase
its energy here on Earth using device B, and store the excited particles in a
battery using device C. Of course, as any other imagination crutch, this one is
personal, and it is not a complete argument.}

To be forced to expend energy every time we want to perform a (given type of)
quantum measurement is of course as physical as anything can get. Since
Einstein (and the famous formula in so many T-shirts), we know that having
energy is equivalent to existing; at least to \emph{physically} existing. This
physical nature of quantum measurements seems absolutely clear from the
previous discussion, but what other kind of existence could measurements have
had? One might want to answer that they could have existed only as
\emph{informational} operations. In fact, that is precisely the way in which
they are assumed to exist in classical mechanics, where, as discussed in
sec.~\ref{sec:introduction}, they can be made subtler and subtler, and no
compulsory minimum energy expenditure to carry them out is required by the
theory. On the contrary, quantum mechanics, it seems, explicitly includes
the---reasonable---requirement that, in order to find out something about a
physical system, we have to probe it in the most real of ways, that is, we have
to prove it \emph{physically}. And what more obvious signature of physicality
that having to expend energy in the process?

This may naturally have some bearing upon a family of approaches to the
foundations of quantum mechanics that are sometimes referred to as
\emph{informational}. Some say the family was founded by Wheeler and, indeed,
his famous motto ``it from bit'' [i.e., physical existence from information
\citep{Wheeler1990}] has become almost a mantra in some circles. The basic idea
is that most (if not all) of the structure of quantum mechanics is related to
information; information about the system for some people, information about
the results of possible human interventions (measurements) on it for others. As
in any general philosophical viewpoint, the positions of the different
researchers come in a wide variety of intensities; from the most radical
[claiming, for example, as Wheeler suggested, that the world \emph{is}
information], to the more nuanced [see, e.g., \citep{Fuchs2001,Fuchs2002b}].

Informational approaches are very powerful in the practical sense, and they
have been indeed instrumental in advancing the most applied quantum fields,
such as computation or cryptography. Moreover, I am confident that any solid
account of the foundations of quantum mechanics has to take them into account
in one way or another. One should not overlook that they seem to deal nicely
(at least at first sight) with some of the most troublesome quantum
``paradoxes'' and the associated conceptual conundrums. However, after noticing
that (at least some) measurements require the expenditure of the same kind of
physical energy that we use to power our transportation and cities, it looks
difficult to argue for the most radical positions that see measurements as
something related \emph{only} to the knowledge that some unspecified human has
about a given experiment. It seems clear that a measurement must be something
that \emph{happens} ``out there'', physically, with or without human
intervention. And yes, maybe the structure of the theory we use to speak about
what happens is related to some kind of information processing and updating.
After all, a great deal of what language is useful for is to process and update
information, and a physical theory is (mostly) made of language; of precise and
sometimes mathematical language, but language in any case. This being said, it
looks a little bit of a wild extrapolation from this---almost
tautological---reflection to suggest that, just because it seems useful to
overlay information theoretic concepts on top of it in order to gain a tighter
control of its doings, the world \emph{is} information, or knowledge, or
judgements, or informed bets [a position termed \emph{informational
immaterialism} in the lucid analysis by \cite{Timpson2010}]. That is, it looks
like a wild extrapolation unless the claim that the world is information means
\emph{just that}: that our best theory about it has an informational form or
flavor (or at least some of its parts have). In such a case the claim might be
true, but only by definition and uninterestingly so.

Finally, let me mention a pair of lines of discussion that have not been tackled
here but may be worth pursuing:

On the one hand, notice that I have chosen simplicity over generality in the
\emph{Gedankenexperiment} in sec.~\ref{sec:gedanken}. Clearly, the structure of
the whole setup can be abstracted and applied not only to the particular case
of a spin-1/2 system with the proposed sequence of measurements, but to the
more general situation of, say, any quantum system that we prepare at $t_0 = 0$
in the ground state of the corresponding Hamiltonian, and on which we perform a
measurement of any operator which does not commute with the said Hamiltonian.
This will inevitably result in the collapse onto a state with non-zero
projection in the subspace orthogonal to the ground-state ray, and thus the
total energy of a large collection of these systems will be increased by the
measurement. It may be interesting to analyze the workings and properties of
this more general case, and possibly also of further generalizations (e.g.,
other preparations at $t_0 = 0$).

On the other hand, it is important to point out that I have circumscribed in
this note to the textbook notion of projective, von Neumann measurements
exclusively. If this constraint is lifted and more general ways of obtaining
information about our quantum system are allowed, it is very likely that
substantial modifications to the whole analysis will have to be made. For
example, one can consider generalized measurements based on positive-operator
valued measures (POVMs) [see, e.g., \citep{Barnett2009,Nielsen2010}], or the
so-called weak measurements \citep{Aharonov1988,Vaidman2009}, or even
non-formalized operations such as asking the person that prepared the state how
she did it. I have shown that (at least some) textbook quantum measurements
require a finite amount of energy to occur, and thus must be regarded as
physical processes. Whether or not the same can be said about other varieties
of measurement is a question for the future.

\section*{Acknowledgements}

\hspace{0.5cm} This work has been supported by the grant Grupo Consolidado 
``Biocomputaci\'on y F\'{\i}sica de Sistemas Complejos'' (DGA, Spain).

\phantomsection
\addcontentsline{toc}{section}{References}

\end{document}